\documentclass[journal,12pt,onecolumn,draftclsnofoot,]{IEEEtran}
\usepackage{amsmath,amsfonts}
\usepackage{algorithmic}
\usepackage{algorithm}
\usepackage{array}
\usepackage[caption=false,font=normalsize,labelfont=sf,textfont=sf]{subfig}
\usepackage{textcomp}
\usepackage{stfloats}
\usepackage{url}
\usepackage{verbatim}
\usepackage{graphicx}
\usepackage{cite}
\usepackage[dvipsnames]{xcolor}

\begin{document}

\title{Energy Efficiency Analysis of Active RIS-enhanced  Wireless Network under Power-Sum Constraint}

\author{Jingdie Xin, Yan Wang, Feng Shu, Feng Zhao, Yifan Zhao, \& Hao Jiang
         \thanks{This work was supported in part by the Hainan Province Science and Technology Special Fund under Grant ZDYF2024GXJS292, and by the National Natural Science Foundation of China under Grant U22A2002, in part by the Scientific Research Fund Project of Hainan University under Grant KYQD(ZR)-21008, in part by the Collaborative Innovation Center of Information Technology, Hainan University, under Grant XTCX2022XXC07, in part by the National Key Research and Development Program of China under Grant 2023YFF0612900. (\textit{Corresponding author: Feng Shu}).}
\thanks{Jingdie Xin, Yan Wang, Feng Shu, and Yifan Zhao are with the School of Information and Communication Engineering, Hainan University, Haikou 570228, China (e-mail: xjingdie@163.com; yanwang@hainanu.edu.cn; shufeng0101@163.com; zyf1001@hainanu.edu.cn).}
\thanks{Feng~Shu is also with the School of Electronic and Optical Engineering, Nanjing University of Science and Technology, Nanjing 210094, China.}
\thanks{Feng~Zhao is with the School of Mechanical and Electrical Engineering, Hainan Vocational University of Science and Technology, Haikou 571126, China, and also with Hainan Provincial Collaborative Innovation Center for Artificial Intelligence Education (e-mail: zf@hvust.edu.cn).}
\thanks{Hao~Jiang is with the National Mobile Communications Research Laboratory, Southeast University, Nanjing 210096, China, and also with the School of Artificial Intelligence, Nanjing University of Information Science and Technology, Nanjing 210044, China (email: jianghao@nuist.edu.cn).}
}

\maketitle

\begin{abstract}
Recently, as a green wireless technology, active reconfigurable intelligent surface (RIS) attracts numerous research activities due to its amplifying ability to combat the double-fading effect compared to passive one. How about its energy efficiency (EE) over passive one? Below, the EE of active RIS-aided wireless network in Rayleigh fading channels is analyzed. Using the law of large numbers, EE is derived as a function of five factors: power allocation factor, the number ($N$) of RIS elements, the total power, the noise variances at RIS and at user. To evaluate each factor's impact, the simple EE function for the concerning factor is given with others fixed. To assess the impact of $N$ on EE, we establish an equation with the EE of active RIS equaling that of passive one, and three methods, bisection, Newton’s method, and simulated annealing, are designed to find the roots of this equation. Simulation results show that as $N$ tends to medium-scale or large-scale, the asymptotic performance formula is consistent with the exact EE expression well. As $N$ varies from small-scale to large-scale, the active RIS intersects passive one at some point ($N_0\approx 2^{10}$). When $N< N_0$, active RIS performs better than passive one in terms of EE. Otherwise, there is a converse conclusion.
\end{abstract}

\begin{IEEEkeywords}
Active reconfigurable intelligent surface, performance analysis, energy efficiency, the law of large numbers, power allocation.
\end{IEEEkeywords}

\section{Introduction}
\label{sec1}

\IEEEPARstart{P}{erformance} analysis of wireless network is the foundation for the design and optimization of communication networks. By evaluating main performance indicators including spectrum efficiency, energy efficiency (EE), and transmission performance loss, it can provide scientific decision-making basis in practical network deployment. In such systems, quantization errors (QE) generated by finite bit radio frequency phase shifters can significantly degrade the signal-to-noise ratio (SNR) at receiver \cite{part0_1}. To mitigate these drawback, a hybrid directional modulation schemes was proposed in \cite{part0_2}, which effectively reduced system performance loss by optimizing the configuration of hybrid-precision phase shifter.

\par As communication technology advances, the sixth-generation (6G) networks are anticipated to achieve revolutionary improvements over fifth-generation (5G) system \cite{ref1}. Nevertheless, 6G still faces critical challenges such as signal transmission loss, multipath fading, and EE constraints \cite{2}. To address these challenges, reconfigurable intelligent surface (RIS) technology has become an effective solution, which offers new opportunities for enhancing signal propagation and network performance \cite{3}, \cite{4}, \cite{part1_1}, \cite{part1_2}.

\par As a key enabling technology for 6G communication networks, passive RIS employs massive arrays of programmable reflector, where every individual element can independently modify the phase of incoming electromagnetic waves \cite{part2_5}. Leveraging this unique beamforming capability, passive RIS demonstrates strong potential in strengthening wireless communication performance. Specifically, when the direct path from the base station (BS) to user is obstructed, RIS can establish an alternative virtual direct path through intelligent signal reflection, thereby significantly improving  system coverage and reliability \cite{part2_4}.

\par In terms of RIS technology application, the authors in \cite{ref3} innovatively proposed an optimal relay selection algorithm based on RIS. This algorithm achieved collaborative optimization of signal strength and transmission delay by simultaneously selecting the optimal relay node and configuring the RIS reflection coefficients. However, practical deployment of such systems faced hardware implementation challenges. In response to this critical issues, a theoretical analysis framework for analyzing RIS phase QE was established \cite{ref5}, and the performance loss at different quantization accuracies were systematically evaluated. The impact of RIS components quantity and SNR on confidentiality performance was analyzed in \cite{new5_31}. Furthermore, in \cite{part2_15}, the RIS-assisted relay system offered considerable advantages to achieve better cost and energy performance than conventional relays, owing to their capability for intelligent signal control.

\par In addition, the authors in \cite{ref6} explored the RIS-enabled system of cell-free multiple-input multiple-output and proposed a joint optimization method that simultaneously optimized the phase shift matrix of RIS and beamforming strategy. Significantly, the application scope of RIS technology continued to expand. To overcome the inherent single stream transmission constraints of traditional directional modulation systems, a dual-stream secure communication framework utilizing intelligent reflective surface (IRS) was introduced in \cite{part2_17}.
Meanwhile, the authors of \cite{part2_11} applied RIS to terahertz communication systems, to address the problem of severe propagation loss characteristic of high-frequency band. These breakthroughs collectively propelled RIS technology from theoretical research to practical implementation, providing groundwork for developing power-efficient and dependable network architectures \cite{part2_12}, \cite{part2_13}.

\par Although passive RIS can optimize wireless channel environment through intelligent reflection, its signal processing capability is limited by its passive characteristics, making it unable to perform active amplification and processing of signals. In contrast, active RIS significantly enhanced the signal reception, processing, and forwarding capabilities by integrating amplification circuits, which provided a more flexible wireless environment control method for 6G communication networks \cite{part3_2}, \cite{11}. In \cite{part3_7}, an active IRS-assisted simultaneous wireless information and power transfer system was used to effectively solve the ``dual fading" problem of passive IRS by joint optimization of BS beamforming and IRS phase shift configuration. To enhance physical layer security, a closed-form power allocation (PA) scheme that optimized the secrecy rate (SR) of secure precoding system was proposed in \cite{ref7}. Building on this, the authors of \cite{part3_4} achieved the coordinated optimization of beamforming and PA through two SR-oriented optimization methods in an active RIS-assisted directional modulation system.

\par Under the IRS-enhanced spatial modulation framework, the authors of \cite{part3_6} proposed three beamforming algorithms and two PA strategies by establishing two simplified channel models to enhance the confidentiality performance of network. Furthermore, a comprehensive performance evaluation of active RIS-assisted wireless network was examined in \cite{ref10},
deriving the asymptotic expression of received SNR of user, and analyzing the impact of phase QE on system performance. Additionally, \cite{part3_0} presented a performance comparison between active and passive RIS-assisted networks. Within the given total power constraint, the optimal allocation ratio of BS transmission power and reflection power of RIS were derived, providing valuable guidance for practical deployment.

\par RIS technology had demonstrated great potential in improving system EE by dynamically regulating electromagnetic wave characteristics \cite{part4_0}. The authors in \cite{part4_3} employed stochastic geometry methods to establish a correlation model between coverage probability and EE in scenarios where direct transmission links were unavailable. Furthermore, the deep reinforcement learning algorithm was proposed in \cite{part4_2} achieved an optimal balance between spectral efficiency and EE through coordinated optimization of RIS reflection parameters and BS precoding matrices.

\par As a green low-power consumption wireless reflecting technique, RIS has attracted a huge amount of research activities from both academia and industry world in the last five years. RIS are also divided into two classes: active and passive. Usually, the former is more energy-efficiency than the latter when the number $N$ of RIS elements tends to small-scale or medium-scale. The natural problem is what is the approximate threshold value $N_0$ of the number $N$ of RIS elements such that the former is equal to the latter in terms of EE. Once the threshold value is determined, then if $N$ is smaller than the threshold $N_0$, the former has a higher EE than the latter. In this paper, we will make a complete investigation of the EE of active RIS-enhanced wireless network, and present a fair comparison with passive RIS under the total power sum constraint to find such a threshold value.

The contributions made in this paper are as follows:

\begin{enumerate}
\item
To analyze the dominant impact factors of EE in an active RIS-assisted wireless network in Rayleigh fading channels, the asymptotic expression of EE for such a  system is given and its simple approximate expression is derived by using  the weak law of large numbers. In this expression, EE is shown to be a joint function of the number ($N$) of active RIS elements, PA factor ($\beta$), total power of BS and active RIS ($P_t$), the noise at active RIS ($\sigma_r^2$) and the noise at user ($\sigma_u^2$). Simulation results show that as the number of active RIS elements tends to medium-scale or large-scale, the asymptotic performance formula fits the exact EE expression well. In the following, we will make an analysis of each impact factor on EE with other factors being fixed.

\item
Specifically, since the active RIS requires additional power to amplify the signal, it is necessary to design the optimal PA factor to achieve the maximum EE. By performing a first-order Taylor approximation to the EE, a closed-form expression for $\beta$ is derived. Meanwhile, based on the derived EE expression, the analysis shows that, on the one hand, when $P_t$ tends to infinity, the EE converges to 0 due to the effect of the logarithmic function property; on the other hand, when $\sigma_r^2$ and $\sigma_u^2$ go towards infinitesimal, respectively, the EE approaches a constant value. Simulation results further indicate that when $P_t$ exceeds its optimal value, increasing its value further cannot continue to enhance the EE. Furthermore, when $\sigma_r^2$ is fixed, EE decreases with the increase of $\sigma_u^2$. Similarly, when $\sigma_u^2$ remains constant, EE also decreases with the increase of $\sigma_r^2$.

\item 
Different from the passive RIS-aided network in which the EE increases monotonically with $N$, the EE of the active RIS-aided system does not increase infinitely with $N$ due to the introduction of amplified noise. To find $N$ when the EE of active and passive RIS-aided systems is equal, three methods, namely Newton's method, bisection method (BiS), and simulated annealing method, are provided. Simulation results indicate that the EE of active and passive RIS intersect at a certain number $N_0$ of RIS elements ($N_0\approx 2^{10}$). Specifically, the active RIS outperforms the passive RIS in terms of EE when $N$ does not exceed $N_0$. Otherwise, the opposite conclusion is drawn. 
\end{enumerate}

\par The organizational structure is presented as follows. Section \ref{Section2} presents the system model for an active RIS-aided communication network.
Section \ref{Section3} derives the asymptotic expression of EE. In section \ref{Section4}, a closed-form solution for PA factor $\beta$ is obtained, and the influencing factors of EE are analyzed. In section \ref{Section5}, the determination of $N$ for active and passive RIS-assisted wireless networks with equal EE is studied. 
Section \ref{Section6} provides the simulation outcomes, and section \ref{Section7} summarizes this paper.

\par Notations: Using lowercase letters to represent scalars. 
$*$ represents conjugation, $|\cdot|$ represents modulus. The operator $\mathbb{E}\{\cdot\}$ denotes the mathematical expectation.

\section{System Model}
\label{Section2}

\par
\begin{figure}[h]
    \centering
    \includegraphics[width=0.5\linewidth]
    {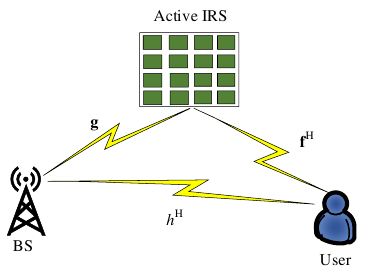}
    \caption{The model of wireless communication system assisted by active RIS.}
    \label{fig:1}
\end{figure}

 An active RIS-assisted wireless network architecture is shown in Fig. $\ref{fig:1}$. This network configuration is composed of a single-antenna BS, a single-antenna user, and an active RIS with $N$ reflection elements.

\par Following the system architecture illustrated in Fig. $\ref{fig:1}$, the signal at the $n$-th active RIS element can be expressed as
\begin{align}\label{activey_{in}(n)}
    y_{in}(n)=\sqrt{\beta P_tL_g}g(n)s+w_r(n),
\end{align}
\noindent where $\beta$ denotes the PA factor between BS and active RIS, $P_t$ represents the total power that includes BS and RIS, where the power of BS and active RIS is $\beta P_t$ and $(1-\beta) P_t$, respectively, $s$ represents the transmission signal of BS. $g(n)=|g(n)|e^{j2\pi\phi_g(n)}$ represents the channel from BS to the $n$-th active RIS element, where $|g(n)|$ represents the amplitude while $\phi_g(n)$ denotes the phase. $L_g$ represents the path loss coefficient of $g(n)$. $w_r(n)$ is the additive white Gaussian noise (AWGN) at the $n$-th active RIS element, following the distribution $w_r(n)\sim \mathcal{CN}(0,\sigma^2_r)$.

\par The signal transmitted from the $n$-th active RIS element is represented by
\begin{align}\label{activey_{out}(n)}
    y_{out}(n)=\sqrt{\beta P_tL_g}p(n)g(n)s+p(n)w_r(n),
\end{align}
\noindent where $p(n)=|p(n)|e^{j\theta(n)}$ represents the amplifying coefficient of the $n$-th active RIS element, while $|p(n)|$ and $\theta(n)$ represent the amplitude gain and phase adjustment, respectively.

\par The signal received by user can be modeled as 
\begin{align}\label{activey^u}
&y_u=\sqrt{\beta P_t}\big(\sqrt{L_fL_g}\sum_{n=1}^{N} f^\ast(n)p(n)g(n)+\sqrt{L_h}h^\ast\big)s+\sqrt{L_f}\sum_{n=1}^{N} f^\ast(n)p(n)w_r(n)+w_u\nonumber\\
&~~~=\sqrt{\beta P_t}\big(\sqrt{L_fL_g}\sum_{n=1}^{N}|f(n)||p(n)||g(n)|\cdot e^{j\big(2\pi\phi_g(n)+\theta(n)-2\pi\phi_f(n)\big)}+\sqrt{L_h}|h|e^{-j\phi_{h}}\big)s\nonumber\\
&~~~~~+\sqrt{L_f}\underbrace{\sum_{n=1}^{N} f^\ast(n)p(n)w_r(n)}_M+w_u\nonumber\\
&~~~=\sqrt{\beta P_t}e^{-j\phi_{h}}\big(\sqrt{L_fL_g}\sum_{n=1}^{N}|f(n)||p(n)||g(n)|\cdot e^{j\big(2\pi\phi_g(n)+\theta(n)-2\pi\phi_f(n)+\phi_{h}\big)}+\sqrt{L_h}|h|\big)s\nonumber\\
&~~~~~+\sqrt{L_f}M+w_u,
\end{align}
\noindent where $f^\ast(n)=|f(n)|e^{-j2\pi\phi_f(n)}$ and $h^\ast=|h|e^{-j\phi_{h}}$ are the channels from $n$-th active RIS element to user and BS to user, respectively, where $|f(n)|$, $|h|$ and $\phi_f(n)$, $\phi_{h}$ denote the amplitude and phase of the corresponding channels. $L_f$ and $L_h$ represent the path loss coefficients of $f^\ast(n)$ and $h^\ast$. $w_u\sim \mathcal{C N}\left(0, \sigma^{2}_u\right)$ is the AWGN at user.

\par Assuming the phase shifter configured in active RIS is continuous, therefore, the signal transmitted by BS can ideally be reflected back to user by active RIS. Thus, the phase shift applied by the $n$-th active RIS element is expressed as
\begin{align}
    \theta(n)=2\pi\phi_f(n)-2\pi\phi_g(n)-\phi_{h},
\end{align}
\noindent to simplify the subsequent analysis, the direct path phase is set as $\phi_{h}=0$. Therefore, $(\ref{activey^u})$ can ultimately be transformed into
\begin{align}\label{activewidetildeyu}
    &\widetilde{y}_u=\sqrt{\beta P_t}\big(\sqrt{L_fL_g}\sum_{n=1}^{N}|f(n)||p(n)||g(n)|+\sqrt{L_h}|h|\big)s+\sqrt{L_f}M+w_u.
\end{align}

\par According to ($\ref{activewidetildeyu}$), the SNR at user can be obtained as 
\begin{align}\label{activegamma1}
    &\gamma_u=\frac{\beta P_t\big(\sqrt{L_fL_g}\sum_{n=1}^{N}|f(n)||p(n)||g(n)|+\sqrt{L_h}|h|\big)^2}{L_f\mathbb{E}\{M^HM\}+\sigma^2_u},
\end{align}
\noindent where
\begin{align}\label{activeeq:12EM}
    &\mathbb{E}\{M^HM\}=\mathbb{E}\Big\{\sum_{n=1}^{N}\sum_{m=1}^{M} f^\ast(n)p(n)w(n)f(m)p^\ast(m)w^\ast(m)\Big\}\nonumber\\
    &~~~~~~~~~~~~~=\sum_{n=1}^{N}\sum_{m=1}^{M} f^\ast(n)p(n)f(m)p^\ast(m)\mathbb{E}\Big\{w(n)w^\ast(m)\Big\},
\end{align}
\noindent with
\begin{align}\label{activeeq:13EM}
    \mathbb{E}\Big\{w(n)w^\ast(m)\Big\}=\sigma^2_r\cdot\delta(m-n),
\end{align}
under the condition of $m \neq n$, it follows that $\delta(m-n)=0$, and thus the simplified expression for ($\ref{activeeq:12EM}$) is as follows
\begin{align}\label{simplifiedM^HM}
    \mathbb{E}\{M^HM\}=\sigma^2_r\sum_{n=1}^{N}|f(n)|^2|p(n)|^2.
\end{align}

\par According to \cite{powerconsumpation}, the derivation of EE requires calculating the total power consumption of the system. Firstly, according to ($\ref{activey_{in}(n)}$), the power received by active RIS can be obtained as
\begin{align}\label{activeP_{in}}
    &P_{in}=\sum_{n=1}^{N}y_{in}^\ast(n)y_{in}(n)=\beta P_tL_g\sum_{n=1}^{N}|g(n)|^2+N\sigma^2_r,
\end{align}
\noindent furthermore, based on ($\ref{activey_{out}(n)}$), the power that reflects from active RIS is represented as
\begin{align}\label{activeP_{out}}
    &P_{out}=(1-\beta)P_t=\sum_{n=1}^{N}y_{out}^\ast(n)y_{out}(n)=\beta P_tL_g\sum_{n=1}^{N}|p(n)g(n)|^2+\sigma^2_r\sum_{n=1}^{N}|p(n)|^2,
\end{align}
\noindent as per \eqref{activeP_{in}} and ($\ref{activeP_{out}}$), the total power consumption of active RIS-assisted system can ultimately be expressed as 
\begin{align}\label{activeP_tot}
    &P_{tot}=P_{out}-P_{in}+\mu \beta P_t+P_c\nonumber\\
    &~~~~~=(1-\beta)P_t-\beta P_tL_g\sum_{n=1}^{N}|g(n)|^2-N\sigma^2_r+\mu \beta P_t+P_c,
\end{align}
\noindent where $\mu$ denotes the inverse of BS transmission amplifier efficiency. $ P_c=P_0+NP_{c,n}+P_{0,RIS}$, where $P_{c,n}$ represents the fixed power required of each RIS component for reconfiguration purposes, and $P_{0,RIS}$ is the sum of all other fixed power consumptions in RIS, $P_0$ represents the static power consumed by nodes other than RIS.

\par Following reference \cite{ref10}, assuming that the active RIS uses the same amplification factor, $(\ref{activeP_{out}})$ is derived as follows
\begin{align}\label{active|p(n)|}
    &|p(n)|=\sqrt{\frac{(1-\beta)P_{t}}{\beta P_tL_g\sum_{n=1}^{N}|g(n)|^2+N\sigma^2_r}}.
\end{align}

\par Finally, the EE of active RIS-assisted communication network can be fully represented as
\begin{align}
    \mathrm{EE}=B\cdot\frac{\mathrm{log}_2(1+\gamma_u)}{P_{tot}},
\end{align}
\noindent where $B$ denotes the bandwidth of the channel.

\section{Derivation of EE Asymptotic Expression}
\label{Section3}

\par In this section, the asymptotic expression of EE for active RIS-assisted system is derived based on the law of large numbers, and the function of EE with respect to $\beta$, $N$, $P_{t}$, $\sigma^2_r$, and $\sigma^2_u$ is derived.

\par Based on the application of the law of large numbers \cite{dashudinglv}, \eqref{active|p(n)|} is convertible to
\begin{align}\label{dashudinglvactive|p(n)|}
    &|p(n)|=\lambda\approx \sqrt{\frac{(1-\beta)P_{t}}{\beta P_tL_gN\cdot \frac{1}{N}\sum_{n=1}^{N}|g(n)|^2+N\sigma^2_r}}=\sqrt{\frac{(1-\beta)P_{t}}{\beta P_tL_gN\mathbb{E}\big\{|g(n)|^2\big\}+N\sigma^2_r}},
\end{align}
\noindent further, substituting \eqref{dashudinglvactive|p(n)|} into \eqref{simplifiedM^HM} yields
\begin{align}\label{dashujinsiM^HM}
    &\mathbb{E}\{M^HM\}=\sigma^2_rN\cdot \frac{1}{N}\sum_{n=1}^{N}|f(n)|^2|p(n)|^2\nonumber\\
    &~~~~~~~~~~~~~\approx \sigma^2_rN\mathbb{E}\big\{|f(n)|^2|p(n)|^2\big\}=\lambda^2\sigma^2_rN\mathbb{E}\big\{|f(n)|^2\big\},
\end{align}
\noindent since $|f(n)|$ and $|g(n)|$ are independent, according to \eqref{dashudinglvactive|p(n)|} and \eqref{dashujinsiM^HM}, the SNR provided by user can be simplified as
\begin{align}\label{dashuactivegammau}
    &\gamma_u=\frac{\beta P_t\big(\lambda\sqrt{L_fL_g}N\frac{1}{N}\sum_{n=1}^{N}|f(n)||g(n)|+\sqrt{L_h}|h|\big)^2}{L_f\lambda^2\sigma^2_rN\mathbb{E}\big\{|f(n)|^2\big\}+\sigma^2_u}\nonumber\\
    &~~~\approx \frac{\beta P_t\big(\lambda\sqrt{L_fL_g}N\mathbb{E}\big\{|f(n)||g(n)|\big\}+\sqrt{L_h}|h|\big)^2}{L_f\lambda^2\sigma^2_rN\mathbb{E}\big\{|f(n)|^2\big\}+\sigma^2_u}\nonumber\\
    &~~~=\frac{\beta P_t\big(\lambda\sqrt{L_fL_g}N\mathbb{E}\big\{|f(n)|\big\}\mathbb{E}\big\{|g(n)|\big\}+\sqrt{L_h}|h|\big)^2}{L_f\lambda^2\sigma^2_rN\mathbb{E}\big\{|f(n)|^2\big\}+\sigma^2_u},
\end{align}
\noindent similarly, \eqref{activeP_tot} can be transformed into
\begin{align}\label{dashuptot}
    &P_{tot}=(1-\beta)P_t-\beta P_tL_gN\cdot\frac{1}{N}\sum_{n=1}^{N}|g(n)|^2-N\sigma^2_r+\mu \beta P_t+P_c\nonumber\\
    &~~~~\approx (1-\beta)P_t-\beta P_tL_gN\mathbb{E}\big\{|g(n)|^2\big\}-N\sigma^2_r+\mu \beta P_t+P_c.
\end{align}

\par Assuming the Rayleigh fading channel is considered, the corresponding probability density function can be expressed as
\begin{align}\label{Rayleigh distribution}
   f_{\alpha}(x)=\left\{\begin{array}{ll}
   \frac{x}{\alpha^{2}} e^{-\frac{x^{2}}{2 \alpha^{2}}}, & x \in[0,+\infty), \\
   0, & \text { otherwise },
   \end{array}\right.
\end{align} 
\noindent with $\alpha >0$ denotes the Rayleigh distribution parameter.

\par Referring to $(\ref{Rayleigh distribution})$, the following expressions can be derived
\begin{align}\label{activeqiwang}
   &\mathbb{E}\big\{|f(n)|\big\}=\big(\frac{\pi}{2}\big)^{\frac{1}{2}}\alpha_f,~~~~~~~~~~\mathbb{E}\big\{|g(n)|\big\}=\big(\frac{\pi}{2}\big)^{\frac{1}{2}}\alpha_g,\nonumber\\
   &\mathbb{E}\big\{|f(n)|^2\big\}=2\alpha_f^2,~~~~~~~~~~~~~~\mathbb{E}\big\{|g(n)|^2\big\}=2\alpha_g^2.
\end{align}

\par According to the above formulas, we can convert $(\ref{dashudinglvactive|p(n)|})$ into
\begin{align}\label{active|p(n)|jinsi}
    &\lambda=\sqrt{\frac{(1-\beta)P_{t}}{\beta P_tL_gN2\alpha^2_g+N\sigma^2_r}},
\end{align}
\noindent by substituting \eqref{activeqiwang} into \eqref{dashuactivegammau}, the SNR at user can be obtained as
\begin{align}\label{activegamma1}
    &\widetilde{\gamma}_u=\frac{\beta P_t}{2NL_f\alpha^2_f\lambda^2\sigma^2_r+\sigma^2_u}(\frac{\pi^2}{4}L_fL_gN^2\lambda^2\alpha^2_f\alpha^2_g+\frac{\pi}{2}L_h\alpha^2_h+\pi(\frac{\pi}{2})^{\frac{1}{2}}\sqrt{L_fL_gL_h}N\lambda\alpha_f\alpha_g\alpha_h),
\end{align}
\noindent the total power consumption of an active RIS-assisted system approximated by the law of large numbers can be expressed as
\begin{align}\label{activePTOT}
    &\widetilde{P}_{tot}=P_t-2N\beta P_tL_g\alpha_g^2-N\sigma^2_r +P_c.
\end{align}

\par Finally, the EE of active RIS-assisted system can be approximated as
\begin{align}
    \mathrm{EE}=B\cdot\frac{\mathrm{log}_2(1+\widetilde{\gamma}_u)}{\widetilde{P}_{tot}},
\end{align}
\noindent the EE can be further simplified as a function of $\beta$, $N$, $P_t$, $\sigma^2_r$, and $\sigma^2_u$, following 
\begin{align}\label{activeGEE1}
		&\mathrm{EE}(\beta, N, P_t, \sigma^2_r,\sigma^2_u)\nonumber\\
        &~~~~~=B\cdot\frac{\mathrm{log}_2\bigg(1+\frac{\beta P_t\Big(A_1NP_t(1-\beta)+A_2\beta P_t+A_3\sigma^2_r+A_4\sqrt{A_5N\beta P^2_t(1-\beta)-N\beta P_t\sigma^2_r+NP_t\sigma^2_r}\Big)}{A_6P_t(1-\beta)\sigma^2_r+A_5\beta P_t\sigma^2_u+\sigma^2_r\sigma^2_u}\bigg)}{-A_5N\beta P_t-N\sigma^2_r+NP_{c,n}+P_t+A_7},
	\end{align}
\noindent where
\begin{align}
    &A_1=\frac{\pi^2}{4}L_gL_f\alpha^2_f\alpha^2_g,~~~~~~~~~~~~~~~~~A_2=\pi L_gL_h\alpha^2_h\alpha^2_g,~~~~~~~~~A_3=\frac{\pi}{2}L_h\alpha^2_h,\nonumber\\
    &A_4=\pi(\frac{\pi}{2})^{\frac{1}{2}}\alpha_g\alpha_f\alpha_h\sqrt{L_fL_gL_h},~~~~A_5=2L_g\alpha^2_g,~~~~~~~~~~~~~~~A_6=2L_f\alpha^2_f,\nonumber\\
    &A_7=P_0+P_{0,RIS}.
\end{align}

\section{Performance Analysis of EE}
\label{Section4}

\par In this section, the main factors influencing the performance of EE, namely $\beta$, $P_{t}$, $\sigma^2_r$, and $\sigma^2_u$ are analyzed. The analytical expression of $\beta$ is solved in subsection $\ref{section4subsection4}$. In subsections $\ref{section4subsection1}$, $\ref{section4subsection2}$, $\ref{section4subsection3}$, the affecting of $P_{t}$, $\sigma^2_r$, $\sigma^2_u$ on EE are analyzed.

\subsection{Performance Analysis of EE on $\beta$}
\label{section4subsection4}

\par In order to obtain a closed-form expression of PA factor, $(\ref{activeGEE1})$ can be written as a function of $\beta$ as follows
\begin{align}
    \mathrm{EE}(\beta)=\frac{B\cdot\mathrm{log}_2(1+\frac{q_1\beta^2+q_2\beta+q_3\beta\sqrt{q_4\beta^2+q_5\beta+q_6}}{q_7\beta+q_8})}{q_4\beta+q_{9}},
\end{align}
\noindent where
\begin{align}
    &q_1=-A_1NP^2_t+A_2P^2_t,~~~~~~q_2=A_1NP^2_t+A_3P_t\sigma^2_r,~~~~~~~~q_3=A_4P^2_t,\nonumber\\
&q_4=-A_5NP_t,~~~~~~~~~~~~~~~~q_5=A_5NP^2_t-NP_t\sigma^2_r,~~~~~~~~~q_6=NP_t\sigma^2_r,\nonumber\\
    &q_7=A_{5}P_t\sigma^2_u-A_{6}P_t\sigma^2_r,~~~~~q_8=A_6P_t\sigma^2_r+\sigma^2_r\sigma^2_u,~~~~~q_{9}=-N\sigma^2_r+NP_{c,n}+P_t+A_7.
\end{align}

\par We begin by defining $\Delta x=\frac{q_4}{q_6}\beta^2+\frac{q_5}{q_6}\beta$, and apply the Taylor expansion to $\sqrt{q_4\beta^2+q_5\beta+q_6}$ when $\Delta x\to 0$. Thereby, we can obtain the following
\begin{align}
    &\sqrt{q_6}\sqrt{1+\frac{q_4}{q_6}\beta^2+\frac{q_5}{q_6}\beta}=\sqrt{q_6}\sqrt{1+\Delta x}\approx\sqrt{q_6}(1+\frac{1}{2}\Delta x)=1+\frac{1}{2}(\frac{q_4}{q_6}\beta^2+\frac{q_5}{q_6}\beta).
\end{align}
\par When $\gamma(\beta)\to0$, the first-order Taylor expansion of $\mathrm{log}_2(1+\gamma(\beta))$ is given as
\begin{align}
    &\mathrm{log}_2(1+\gamma(\beta))\approx\frac{1}{ln2}\gamma(\beta)=\frac{1}{ln2}\frac{\frac{q_3q_4}{2\sqrt{q_6}}\beta^3+(q_1+\frac{q_3q_5}{2\sqrt{q_6}})\beta^2+(q_2+q_3q_6)\beta}{q_7\beta+q_8}.
\end{align}

\par Therefore, $\mathrm{EE}(\beta)$ can ultimately be expressed as
\begin{align}\label{daoshubeta}
   &\widetilde{\mathrm{EE}}(\beta)=\frac{B}{ln2}\cdot\frac{c_1\beta^3+c_2\beta^2+c_3\beta}{c_4\beta^2+c_5\beta+c_6}, 
\end{align}
\noindent where
\begin{align}
     &c_1=\frac{q_3q_4}{2\sqrt{q_6}},~~~~~~c_2=q_1+\frac{q_3q_5}{2\sqrt{q_6}},~~~~~~~~c_3=q_2+q_3q_6,\nonumber\\
     &c_4=q_7q_4,~~~~~~~~c_5=q_7q_{9}+q_8q_4,~~~~~~~c_6=q_8q_{9}.
\end{align}

\par The derivative of ($\ref{daoshubeta}$) with respect to $\beta$ yields
\begin{align}\label{derivativeGEE}
   \widetilde{\mathrm{EE^{'}}}(\beta)=\frac{B}{ln2}\frac{l_1\beta^4+l_2\beta^3+l_3\beta^2+l_4\beta+l_5}{(c_4\beta^2+c_5\beta+c_6)^2}, 
\end{align}
\noindent where
\begin{align}
    &l_1=c_1c_4,~~~~~~~l_2=2c_1c_5,~~~~~~l_3=3c_1c_6+c_2c_5-c_3c_4,\nonumber\\
    &l_4=2c_2c_6,~~~~~~~~~~~~~~~~~~~~~~~~~l_5=c_3c_6.
\end{align}

\par Letting ($\ref{derivativeGEE}$) equal zero, following Ferrari’s method \cite{ref16}, the four candidate solutions are represented as
\begin{align}
   \hat{\beta}_{1,2} =-\frac{E_1}{4}+\frac{\eta_1}{2}\pm\frac{\mu_1}{2},~~~\hat{\beta}_{3,4} =-\frac{E_1}{4}-\frac{\eta_1}{2}\pm\frac{\mu_2}{2},
\end{align}
\noindent where
\begin{align}
    &\eta_1=\sqrt{\frac{E^2_1}{4}-B_1+\gamma_1},~~~~~\alpha_1=\frac{1}{3}(3E_1C_1-12D_1-B^2_1),\nonumber\\
    &\gamma_1=\frac{B_1}{3}+\sqrt[3]{-\frac{\beta_1}{2}+\sqrt{\frac{\beta^2_1}{4}+\frac{\alpha^3_1}{27}}}+\sqrt[3]{-\frac{\beta_1}{2}-\sqrt{\frac{\beta^2_1}{4}+\frac{\alpha^3_1}{27}}},\nonumber\\
    &\beta_1=\frac{1}{27}(-2B^3_1+9E_1B_1C_1+72B_1D_1-27C^2_1-27E^2_1D_1),\nonumber\\
    &\alpha_1=\frac{1}{3}(3E_1C_1-12D_1-B^2_1),\nonumber\\
    &\mu_1=\sqrt{\frac{3}{4}E^2_1-\eta^2_1-2B_1+\frac{1}{4\eta_1}(4E_1B_1-8C_1-E^3_1)},\nonumber\\
    &\mu_2=\sqrt{\frac{3}{4}E^2_1-\eta^2_1-2B_1-\frac{1}{4\eta_1}(4E_1B_1-8C_1-E^3_1)},\nonumber\\
    &E_1=\frac{l_2}{l_1},~~~~~B_1=\frac{l_3}{l_1},~~~~~C_1=\frac{l_4}{l_1},~~~~~D_1=\frac{l_5}{l_1}.
\end{align}

\par Considering that $\beta$ is limited to the interval of $[0, 1]$, we need to perform feasibility analysis on the four candidate solutions mentioned above, and then construct new forms of the four candidate solutions as follows
\begin{align}\label{eq:14}
    \widetilde{\beta}_i=\left\{\begin{array}{ll}
    \hat{\beta}_i, & \hat{\beta} \in[0,1], \\
    0,& \hat{\beta} \notin[0,1],
    \end{array}\right.
\end{align} 
\noindent where $i=1, 2, 3, 4$, the optimal candidate solution follows
\begin{align}
    \beta_{ b}=\underset{\beta \in S_{b}}{\operatorname{argmax}}\{\mathrm{EE}(\beta)\},
\end{align}
\noindent where
\begin{align}\label{S_b}
   S_b=\{0,1,\widetilde{\beta}_1,\widetilde{\beta}_2,\widetilde{\beta}_3,\widetilde{\beta}_4\}. 
\end{align}

\subsection{Performance Analysis of EE on $P_t$}
\label{section4subsection1}

\par To analyze the influence of $P_{t}$ on $\mathrm{EE}$, $(\ref{activeGEE1})$ is reformulated as a function of $P_{t}$ as follows
\begin{align}\label{GEEPt}
    \mathrm{EE}(P_t)=\frac{B\cdot\mathrm{log}_2(1+\frac{k_1P^2_t+k_2P_t+k_3P_t\sqrt{k_4P^2_t+k_5P_t}}{k_6P_t+k_7})}{k_8P_t+k_9},
\end{align}
\noindent where
\begin{align}
    &k_1=A_1N\beta (1-\beta)+A_2\beta^2, ~~~~k_2=A_3\beta \sigma^2_r,~~~~~~~~~~~~~~~k_3=A_4\beta,\nonumber\\
    &k_4=A_5N\beta (1-\beta),~~~~~~~~~~k_5=-N\beta \sigma^2_r+N\sigma^2_r,~~~~~~~~k_6=A_6(1-\beta)\sigma^2_r+A_5\beta \sigma^2_u,\nonumber\\
    &k_7=\sigma^2_r\sigma^2_u,~~~~~~~~~~~~~~~~~~~~k_8=-A_{5}N\beta+1,~~~~~~~~~~~~k_9=-N\sigma^2_r+NP_{c,n}+A_{7},
\end{align}
\noindent when $P_t\to\infty$, the expression simplifies to
\begin{align}\label{Pttoinfty}
    &\mathrm{EE}\to\frac{B\cdot\mathrm{log}_2(1+\frac{A_1N\beta (1-\beta)+A_2\beta^2+A_4\beta\sqrt{A_5N\beta (1-\beta)}}{A_6(1-\beta)\sigma^2_r+A_5\beta \sigma^2_u}P_t)}{(-A_{5}N\beta+1)P_t},
\end{align}
\noindent from the above expression, it can be seen that with other parameters fixed, EE is a nonlinear function with respect to $P_t$. Since the denominator of EE is a linear function with respect to $P_t$ and the numerator is a logarithmic function of $P_t$, when $P_t$ approaches infinity ($P_t\to\infty$), EE tends to 0. When $P_t$ is fixed, EE shows a decreasing trend with the increase of $N$, which is due to the fact that the denominator grows linearly with $N$ while the numerator grows at a slower rate than the linear growth.

\subsection{Performance Analysis of EE on $\sigma_r^2$}
\label{section4subsection2}

\par Due to the introduction of noise by active RIS, it is necessary to examine the impact of $\sigma^2_r$ on EE. $(\ref{activeGEE1})$ can be expressed as
\begin{align}\label{GEEsigmaR}
    &\mathrm{EE}(\sigma^2_r)=\frac{B\cdot\mathrm{log}_2(1+\frac{a_1\sigma^2_r+a_2\sqrt{a_3\sigma^2_r+a_4}+a_5}{a_6\sigma^2_r+a_7})}{-N\sigma^2_r+a_8},
\end{align}
\noindent where
\begin{align}
   & a_1=A_3\beta P_t,~~~~~~~~~~~~~~~~~~a_2=A_4\beta P_t,~~~~~~~~~~~~~~~~~~~~~~~~~~~~a_3=-N\beta P_t+N P_t,\nonumber\\
   &a_4=A_5N\beta P^2_t(1-\beta),~~~~~~a_5=A_1N\beta P^2_t(1-\beta)+A_2\beta^2P^2_t,~~~a_6=A_6P_t(1-\beta)+\sigma^2_u,\nonumber\\
   &a_7=A_{5}\beta P_t\sigma^2_u,~~~~~~~~~~~~~~~a_8=-A_{5}N\beta P_t+NP_{c,n}+P_t+A_{7},
\end{align}
\noindent when $\sigma^2_r\to 0$, then (\ref{GEEsigmaR}) is simplified as \begin{align}\label{sigmarGEEquxiang}
		 &\mathrm{EE}\to B\cdot\frac{\mathrm{log}_2\Big(1+\frac{\beta P_t\big(A_1NP_t(1-\beta)+A_2\beta P_t+A_4\sqrt{A_5N\beta P^2_t(1-\beta)}\big)}{A_{5}\beta P_t\sigma^2_u}\Big)}{-A_{5}N\beta P_t+NP_{c,n}+P_t+A_{7}},
	\end{align}

\noindent by analyzing the expression, it can be concluded that when other parameters of the system are fixed, EE converges to a certain value. Additionally, it can be seen from the expression that EE decreases with the increase of $\sigma^2_u$, because the increase of $\sigma^2_u$ leads to a decrease in SNR, which in turn leads to a decrease in EE.

\subsection{Performance Analysis of EE on $\sigma_u^2$}
\label{section4subsection3}

\par Following the same approach, we study the effect of $\sigma^2_u$ on EE by rephrasing the expression for $\sigma^2_u$ in ($\ref{activeGEE1}$) as follows
\begin{align}\label{GEEsigmaU}
    &\mathrm{EE}(\sigma^2_u)=\frac{B}{d_1}\cdot\mathrm{log}_2(1+\frac{d_2}{d_3\sigma^2_u+d_4}),
\end{align}
\noindent where
\begin{align}
    & d_1=-A_{5}N\beta P_t-N\sigma^2_r+NP_{c,n}+P_t+A_7,\nonumber\\
   & d_2=\beta P_t\Big(A_1NP_t(1-\beta)+A_2\beta P_t+A_3\sigma^2_r+A_4\sqrt{A_5N\beta P^2_t(1-\beta)-N\beta P_t\sigma^2_r+NP_t\sigma^2_r}\Big),\nonumber\\
   & d_3=A_{5}\beta P_t+\sigma^2_r,~~~~~~ d_4=A_6P_t(1-\beta)\sigma^2_r,
\end{align}
\noindent when $\sigma^2_u\to 0$, then we can obtain the following
\begin{align}\label{GEEsigmauquxiang}
     &\mathrm{EE}\to B\cdot \frac{\mathrm{log}_2\Big(1+\frac{\beta P_t\big(A_1NP_t(1-\beta)+A_2\beta P_t+A_3\sigma^2_r+A_4\sqrt{A_5N\beta P^2_t(1-\beta)-N\beta P_t\sigma^2_r+NP_t\sigma^2_r}\big)}{A_6P_t(1-\beta)\sigma^2_r}\Big)}{-A_{5}N\beta P_t-N\sigma^2_r+NP_{c,n}+P_t+A_7},
\end{align}
\noindent similarly, when other parameters are given, $\sigma^2_u$ approaching 0 ($\sigma^2_u\to0$) causes EE to converge to a constant value. Furthermore, as $\sigma^2_r$ in \eqref{GEEsigmauquxiang} increases, the total power consumption of the system increases linearly. However, due to the significantly lower growth rate of SNR compared to power consumption, EE shows a downward trend with the increase of $\sigma^2_r$.

\section{Relationship of EE Between Active and Passive RIS-aided Networks}
\label{Section5}

\par In this section, the determination of $N$ is studied when the EE of active and passive RIS-assisted systems is equal. By defining $\alpha=\frac{1}{N}$, the expressions of EE for active and passive RIS-assisted systems with respect to $\alpha$ are derived, and based on the condition that their EE is equal, $f(\alpha)$ is finally obtained. 

\par Regarding the determination of $\alpha$, three solution algorithms, namely Newton's method, BiS and simulated annealing method, are proposed in subsections $\ref{Section3subsection1}$, $\ref{Section3subsection2}$, and $\ref{Section3subsection3}$, respectively.

\par The EE function of the active RIS assisted system with respect to $N$ can be given by
\begin{align}
    \mathrm{EE}^a(N)=\frac{B\cdot\mathrm{log}_2(1+m_1N+m_2\sqrt{N}+m_3)}{m_4N+m_{5}},
\end{align}
\noindent where
\begin{align}
    &m_1=\frac{A_1P^2_t \beta (1-\beta)}{A_6P_t(1-\beta)\sigma^2_r+A_5\beta P_t\sigma^2_u+\sigma^2_r\sigma^2_u},~~~~~~~~~~~m_2=\frac{A_4\sqrt{A_5\beta P^2_t(1-\beta)-\beta P_t\sigma^2_r+P_t\sigma^2_r}}{A_6P_t(1-\beta)\sigma^2_r+A_5\beta P_t\sigma^2_u+\sigma^2_r\sigma^2_u},\nonumber\\
    &m_3=\frac{A_2\beta^2P^2_t+A_3\beta P_t\sigma^2_r}{A_6P_t(1-\beta)\sigma^2_r+A_5\beta P_t\sigma^2_u+\sigma^2_r\sigma^2_u},~~~~~~~~~~~m_4=-A_5\beta P_t-\sigma^2_r+P_{c,n},\nonumber\\
    &m_{5}=P_t+A_7.
\end{align}

\par Similar to the derivation of EE for active RIS-assisted system, the EE of passive RIS-assisted system can be represented as
\begin{align}\label{passiveEE}
    \mathrm{EE}^p(N, P_t, \sigma^2_u)=B\cdot\frac{\mathrm{log}_2(1+\frac{P_t(A_8N^2+A_9N+A_{10}}{\sigma^2_u})}{P_t+NP^p_{c,n}+A_{11}},
\end{align}
\noindent where
\begin{align}
    &A_8=\frac{\pi^2}{4}L_fL_g\alpha^2_f\alpha^2_g,~~A_9=\pi(\frac{\pi}{2})^{\frac{1}{2}}\sqrt{L_fL_gL_h}\alpha_f\alpha_g\alpha_h,~~~A_{10}=\frac{\pi}{2}L_h\alpha^2_h,~~A_{11}=P_0+P^p_{0,RIS},
\end{align}
\noindent the specific derivation process can be found in the \textbf{Appendix}.

\par According to ($\ref{passiveEE}$), $\mathrm{EE}^p$ can be expressed as a function of $N$, as follows
\begin{align}\label{passiveGEEN}
    \mathrm{EE}^p(N)=\frac{B\cdot\mathrm{log}_2(1+m_6N^2+m_7N+m_8)}{P^p_{c,n}N+m_{9}},
\end{align}
\noindent where
\begin{align}
    &m_6=\frac{\pi^2}{4}\cdot \frac{1}{\sigma^2_u}P_tL_fL_g\alpha^2_f\alpha^2_g,~~~~~~~m_7=\pi (\frac{\pi}{2})^\frac{1}{2}\frac{1}{\sigma^2_u} P_t\sqrt{L_fL_gL_h}\alpha_f\alpha_g\alpha_h,\nonumber\\
    &m_8=\frac{\pi}{2}\frac{1}{\sigma^2_u}P_tL_h\alpha^2_h,~~~~~~~~~~~~~~~~m_{9}=P_t+P_0+P^p_{0,RIS}.
\end{align}

\par To determine $N$ when the EE of the active and passive RIS-assisted systems is equal, we have
\begin{align}\label{GEEaGEEp}
    \mathrm{EE}^a(N)=\mathrm{EE}^p(N),
\end{align}
\noindent as a consequence, the derived expression is as follows
\begin{align}\label{GEEaGEEp}
    &\frac{\mathrm{log}_2(1+m_1N+m_2\sqrt{N}+m_3)}{m_4N+m_{5}}=\frac{\mathrm{log}_2(1+m_6N^2+m_7N+m_8)}{P^p_{c,n}N+m_{9}}. 
\end{align}

\par When defined $\alpha$ = $\frac{1}{{N}}\in (0,1]$,  ($\ref{GEEaGEEp}$) can be further rewritten as
\begin{align}
    f(\alpha)=f_1(\alpha)-f_2(\alpha)=0,
\end{align}
\noindent where
\begin{align}
    &f_1(\alpha)=\frac{\mathrm{log}_2(1+\frac{m_1+m_2\sqrt{\alpha}+m_3\alpha}{\alpha})}{m_4+m_{5}\alpha},~~~~~~~f_2(\alpha)=\frac{\mathrm{log}_2(1+\frac{m_6+m_7\alpha+m_8\alpha^2}{\alpha^2})}{P^p_{c,n}+m_{9}\alpha}.
\end{align}

\par To solve $\alpha$, three typical algorithms are given below.

\subsection{Newton's method}
\label{Section3subsection1}

\par The Newton's method provides an efficient numerical iterative approach by solving $f(\alpha)=0$ \cite{netown}. The base principle is to achieve fast convergence according to the method of successive local linearization.

\par The iterative update process of $\alpha$ in the $i$-th iteration of Newton's method can be described by the following mathematical expression
\begin{align}
    \alpha^i=\alpha^{i-1}-\frac{f(\alpha^{i-1})}{f'(\alpha^{i-1})},
\end{align}
\noindent where
\begin{align}
    f^{\prime}(\alpha)=f_1'(\alpha)-f_2'(\alpha),
\end{align}
\noindent with
\begin{align}
    &f_1'(\alpha)=\frac{1}{(m_4+m_5\alpha)^2}\Big(\frac{(-m_2\alpha-2m_1)(m_4+m_5\alpha)}{\mathrm{ln2}\cdot ((m_3+1)\alpha^3+m_2\alpha^2+m_1\alpha)}\nonumber\\
    &~~~~~~~~~~-m_5\cdot \mathrm{log_2}(1+\frac{m_1+m_2\alpha+m_3\alpha^2}{\alpha^2})\Big),\nonumber\\
    &f_2'(\alpha)=\frac{1}{(P^p_{c,n}+m_{9}\alpha)^2}\Big(\frac{(-\frac{1}{2}m_7\sqrt{\alpha}-m_6)(P^p_{c,n}+m_{9}\alpha)}{\mathrm{ln2}\cdot \alpha((m_8+1)\alpha+m_7\sqrt{\alpha}+m_6)}\nonumber\\
    &~~~~~~~~~~-m_{9}\cdot \mathrm{log_2}(1+\frac{m_6+m_7\sqrt{\alpha}+m_8\alpha}{\alpha})\Big).
\end{align}

\subsection{BiS}
\label{Section3subsection2}

\par The BiS is a numerical approach that approximates the solution of equation $f(\alpha)=0$ by halving the interval \cite{erfen}. This method gradually narrows down the range of roots by repeatedly binary searching the interval, and ultimately determines the numerical solution of the equation with a specified accuracy. Since $\alpha\in (0,1]$, we can define $\epsilon_1=0$, $\epsilon_2=1$, and take the midpoint of $(0,1]$ as $\alpha_1$, which can be expressed as 
 \begin{align}
   \alpha_1= \frac{\epsilon_1+\epsilon_2}{2}, 
 \end{align}
if $f(\alpha_1)=0$, then $\alpha=\alpha_1$. Otherwise, if the signs of $f(\alpha_1)$ and $f(\epsilon_1)$ are opposite, the interval range of $\alpha$ is $[\epsilon_1, \alpha_1]$. On the contrary, $\alpha\in [\alpha_1, \epsilon_2]$.

\subsection{Simulated annealing method}
\label{Section3subsection3}

\par The simulated annealing algorithm searches for the zero point of a function in the solution space by simulating the physical cooling process \cite{part3_4}. The algorithm execution process is divided into three stages. Firstly, initialize the system parameters, including initial temperature, cooling coefficient, and initial solution. Then, at each temperature state, candidate solutions are generated by randomly perturbing the current solution. Finally, based on the energy difference function \eqref{eta} and probability density function \eqref{P(etaT)}, we decide whether to accept this candidate solution. By repeating the cooling process, the solution of the objective function can be obtained.
\begin{align}\label{eta}
    &\eta=f(\alpha_{i+1})-f(\alpha_{i}),
\end{align}
\begin{align}\label{P(etaT)}
    &P(\eta,T)=e^{-\frac{\eta}{T}},
\end{align}
\noindent where $f(\alpha_{i+1})$ and $f(\alpha_{i})$ represent the energy values at the $i$-th and $i+1$-th iterations, and $\eta$ represents the energy difference. In addition, the probability of accepting a new solution is determined by the Metropolis criterion $P(\eta, T)$.

\section{Simulation Results}
\label{Section6}

\par This section involves the parameter configuration in this paper. The specific settings are shown below. BS, RIS and user locations are set as (0 m, 0 m, 0 m), (150 m, 0 m, 0 m), (100 m, 33 m, 0 m), respectively. The Rayleigh distribution parameters are set to $\alpha^2_f=\alpha^2_g=\alpha^2_h=$0.5. The path loss at distance $d$ is
modeled as $L(d)=\mathrm{PL}_0-10a\mathrm{log_{10}{\frac{\textit{d}}{\textit{$d_0$}}}}$, where $\mathrm{PL}_0$ = -30 dB represents the reference path loss at $d_0 = $ 1 m, while $a$ is the path loss exponent. The inverse of BS transmission amplifier efficiency $\mu = 1$. The exponents of BS$\to$RIS, RIS$\to$User, and BS$\to$User channels are set as 2.3, 2.3, and 3.8, respectively.
\begin{figure}[h]
    \centering
    \includegraphics[width=0.5\linewidth]
    {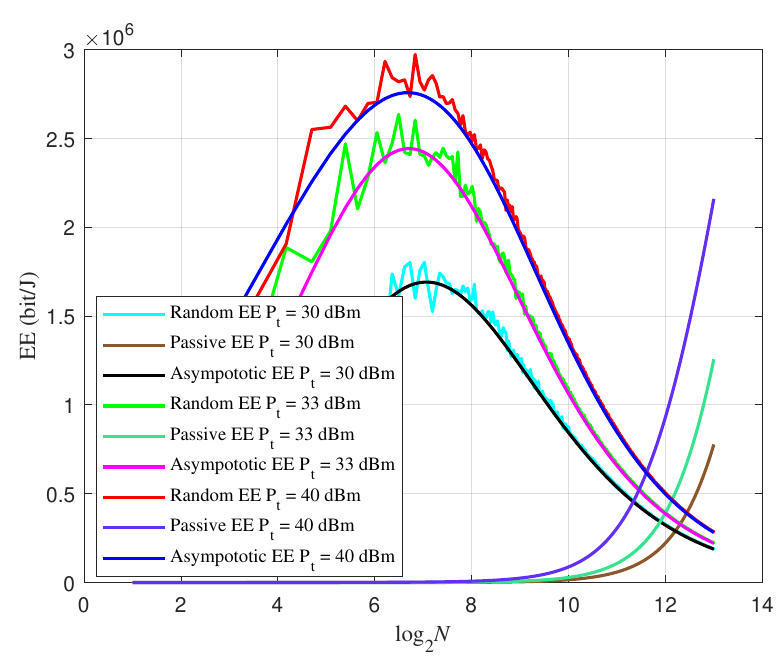}
    \caption{EE versus the number $N$ of RIS element.}
    \label{fig:2}
\end{figure}

\par Fig. $\ref{fig:2}$ presents the curves of EE versus the number $N$ of RIS element. As evidenced by Fig. $\ref{fig:2}$, in the active RIS system, when $N$ tends to the medium-scale or large-scale, the actual EE curve can fit well with the asymptotic EE curve. Due to the linear increase in total power consumption of active RIS with $N$, its EE exhibits a unimodal characteristic of first increasing and then decreasing. As $N$ continues to increase, the passive RIS system ultimately outperforms the active RIS system in terms of EE due to the noise introduced by the active RIS.

\begin{figure}[h]
    \centering
    \includegraphics[width=0.5\linewidth]
    {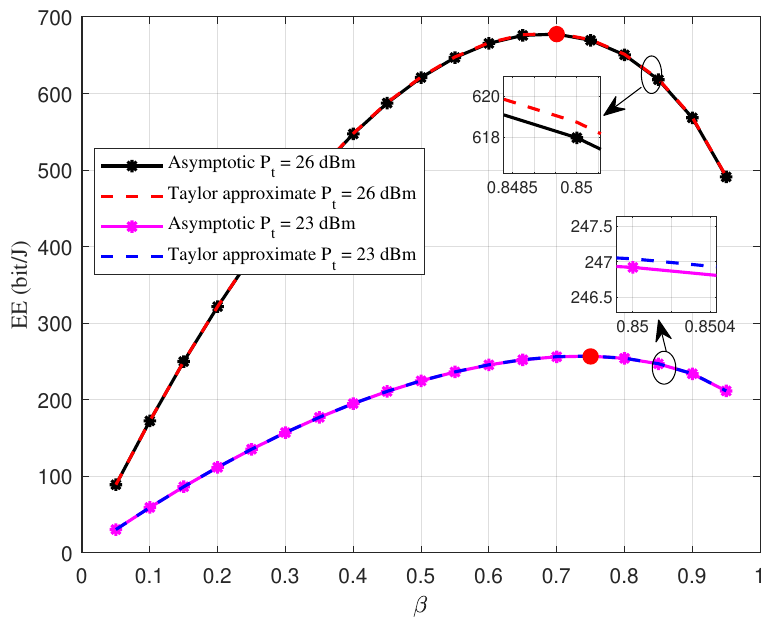}
    \caption{EE versus the PA factor $\beta$.}
    \label{fig:6}
\end{figure}
\par Fig. $\ref{fig:6}$ demonstrates the curves of  EE with its first-order Taylor approximation versus the PA factor $\beta$. From Fig. $\ref{fig:6}$, when PA factor $\beta$ varies between 0 and 1, EE exhibits concave function characteristics, and the first-order Taylor approximation can fit the  EE curve well. Particularly, for different total power $P_t$ configurations, the system has a unique optimal PA factor $\beta$ that maximizes EE, and the candidate solutions at this point are determined in ($\ref{S_b}$).

\begin{figure}[h]
    \centering
    \includegraphics[width=0.5\linewidth]
    {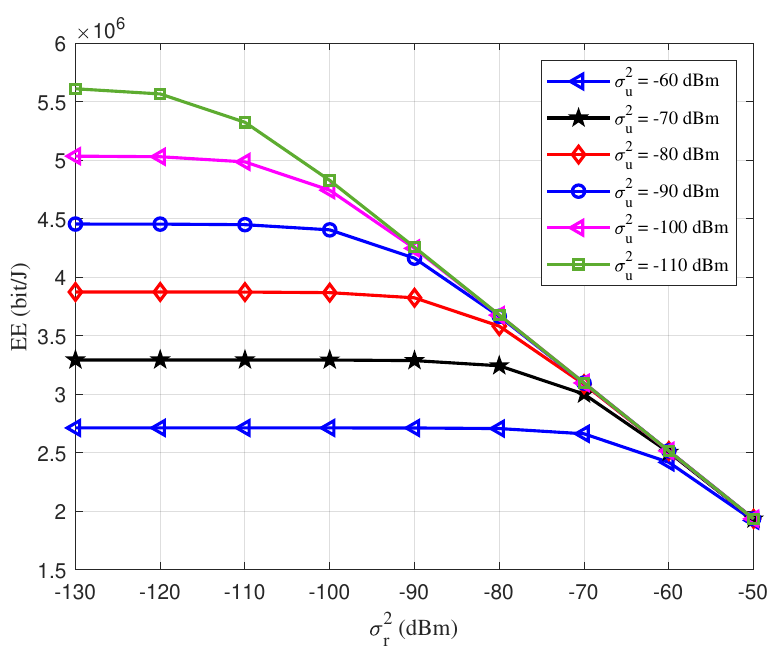}
    \caption{EE versus the noise $\sigma_r^2$ at active RIS.}
    \label{fig:4}
\end{figure}
\par Fig. $\ref{fig:4}$ plots the curves of EE versus  $\sigma^2_r$. It can be seen from Fig. $\ref{fig:4}$ that when $\sigma^2_r$ is small, EE stabilizes near the theoretical analysis value, which is consistent with the theoretical analysis in ($\ref{sigmarGEEquxiang} $). However, as $\sigma^2_r$ further increases, EE gradually shows a decreasing trend. Moreover, when $\sigma^2_r$ is given, increasing  $\sigma^2_u$ will result in a decrease in EE.

\begin{figure}[h]
    \centering
    \includegraphics[width=0.5\linewidth]
    {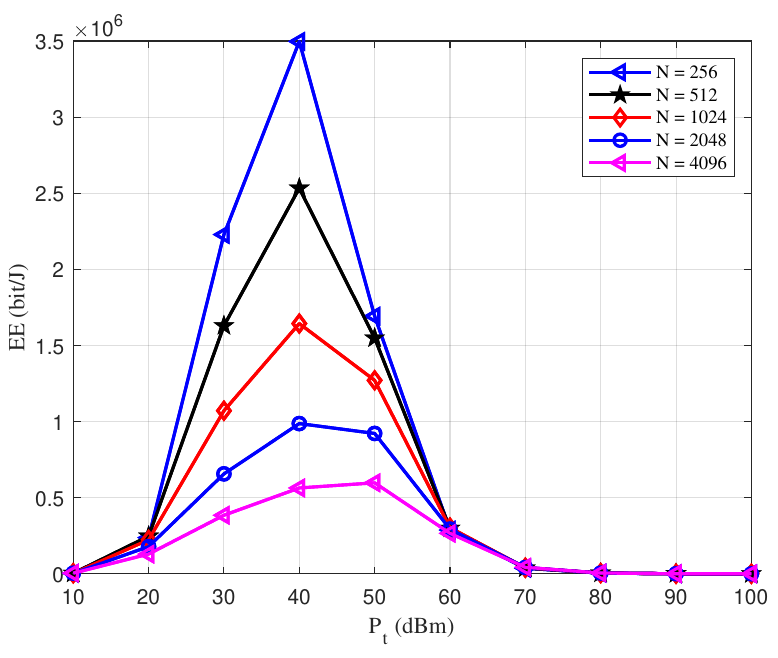}
    \caption{EE versus the total power $P_t$.}
    \label{fig:3}
\end{figure}
\par Fig. $\ref{fig:3}$ illustrates the curves of EE versus the total power $P_t$, where the value of $P_t$ ranges from 10 dBm and 100 dBm. It can be seen that the EE shows a trend of initially rising and then falling, and finally converges 0 in Fig. $\ref{fig:3}$. The reason behind this is that as $P_t$ gradually increases, the total power consumption of the system increases linearly, while the channel capacity exhibits a logarithmic growth characteristic, which leading to EE convergence to zero. For different the numbers of RIS element, there is an optimal $P_t$ for each EE curve, at which point the EE reaches its peak. When $P_t$ exceeds this point, continuing to increase $P_t$ does not improve EE, in exact agreement with the theoretical analysis of \eqref{Pttoinfty}.

\begin{figure}[h]
    \centering
    \includegraphics[width=0.5\linewidth]
    {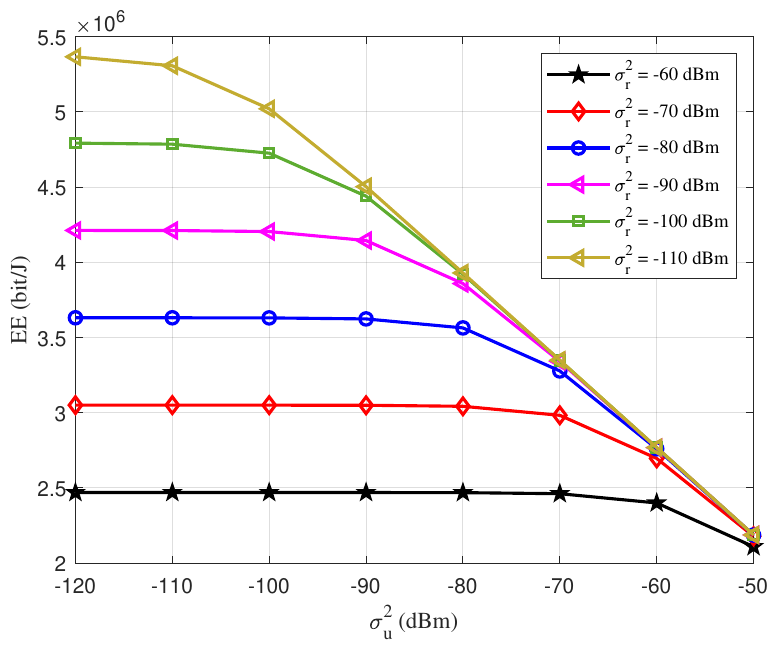}
    \caption{EE versus the noise $\sigma^2_u$ at user.}
    \label{fig:5}
\end{figure}
\par Fig. $\ref{fig:5}$ depicts the curves of EE versus  $\sigma^2_u$. It can be seen that the trend in Fig. $\ref{fig:5}$ is similar to that Fig. $\ref{fig:4}$, i.e., when $\sigma^2_u$ is small, EE tends to a stable value, which matches the analysis results of ($\ref{GEEsigmauquxiang}$). Meanwhile, in Fig. $\ref{fig:5}$, under fixed condition of $\sigma^2_u$, an increase in $\sigma^2_r$ leads to a decrease in EE.

\begin{figure}[h]
    \centering
    \includegraphics[width=0.5\linewidth]
    {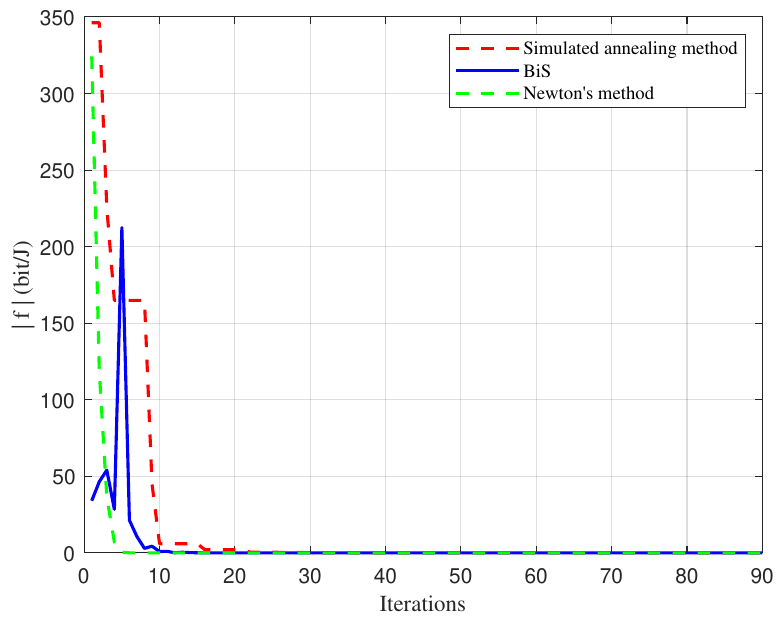}
    \caption{The absolute value of the difference in EE between active and passive RIS-assisted systems $|f|$ versus iterations of simulated annealing method, BiS, and Newton's method.}
    \label{fig:7}
\end{figure}
\par Fig. $\ref{fig:7}$ describes  the convergence curves of three different algorithms, where $|f|$ represents the absolute value of the difference in EE between active and passive RIS-assisted systems. This figure verifies the convergence of Newton's method, simulated annealing method, and BiS after some iterations. It is observed that in terms of the convergence speed, their order is Newton's method $>$ BiS $>$ simulated annealing method, in Fig. $\ref{fig:7}$.

\par
\begin{figure}
    \centering
    \includegraphics[width=0.5\linewidth]
    {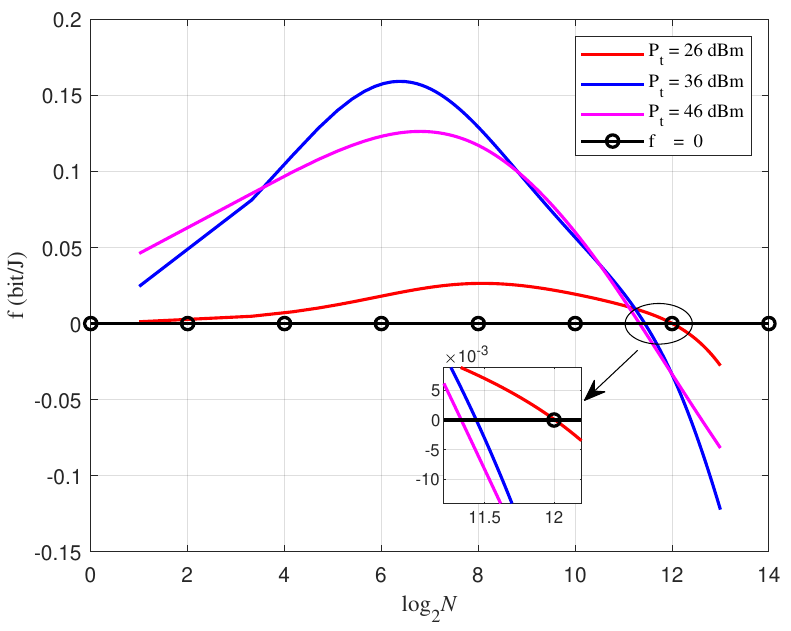}
    \caption{The difference of the EE both active and passive RIS-assisted systems $f$ versus $N$.}
    \label{fig:8}
\end{figure}

 Fig. $\ref{fig:8}$ plots the curves of $f$ versus the $N$, where $f$ is the difference of the EE both active and passive RIS-assisted systems. Under the condition of $\sigma^2_r=\sigma^2_u= -$70 dBm, it can be seen that at different total power levels, each curve has an intersection point with $f_0=0$, which is the point where the EE of the active and passive RIS-assisted systems is equal ($N_0 \approx 2 ^ {10} $). The advantage of active RIS is excellent in terms of EE when $N$ does not exceed $N_0$. Otherwise, the EE performance of passive RIS is better.

\section{Conclusion}
\label{Section7}

\par In this paper, the EE analysis of active RIS-enhanced wireless network under power-sum constraint has been studied. Using the law of large numbers, we derive the asymptotic expression of EE for an active RIS-assisted system, and the expression of EE is shown to be a joint function of $\beta$, $N$, $P_t$, $\sigma^2_r$, and $\sigma^2_u$. In addition, in order to allocate $P_t$ reasonably and maximize EE, by performing the first-order Taylor expansion on EE, a closed-form expression for the $\beta$ is obtained. Theoretical analysis shows that when $P_t$ approaches infinity, EE is asymptotically close to 0. When $\sigma^2_r $ with a fixed $\sigma^2_u$ goes to infinitesimal, EE converges to a constant value. To find the root of equation of the EE of active RIS equalling that of passive one with the number of $N$ as variable, three algorithms, namely BiS, simulated annealing method, and Newton's method, are provided. Simulation results indicate that when $N$ tends to medium-scale and large-scale, the asymptotic performance of EE is better. In particular, when the number of RIS elements reaches the critical value $(N_0\approx 2^{10})$, the EEs of active and passive RIS are equal. When $N$ is less than $N_0$, the EE performance of active RIS is better than that of passive one. Otherwise, the former is worse than the latter in terms of EE.

{\appendix
\par The received signal of the $n$-th RIS element follows
\begin{align}\label{passiveyin}
 y^p_{in}=\sqrt{P_tL_g} g(n)s.  
\end{align}

\par The representation of the reflected signal of the $n$-th RIS element is as follows
\begin{align}\label{passiveyout}
   y^p_{out}=\sqrt{P_tL_g} g(n)p(n)s.  
\end{align}

\par The received signal at user can be represented as
\begin{align}\label{passiveyu}
    &y^p_u=\sqrt{P_t}\big(\sqrt{L_fL_g}\sum_{n=1}^{N} f^\ast(n)p(n)g(n)+\sqrt{L_h}h^\ast\big)s+w_u\nonumber\\
    &~~~=\sqrt{P_t}e^{-j\phi_{h}}\big(\sqrt{L_fL_g}\sum_{n=1}^{N}|f(n)||p(n)||g(n)|\cdot e^{j\big(2\pi\phi_g(n)+\theta(n)-2\pi\phi_f(n)+\phi_{h}\big)}+\sqrt{L_h}|h|\big)s+w_u,
\end{align}
\noindent assuming the phase of the direct channel is $\phi_{h}=0$, similar to the phase alignment method of active RIS, the received signal at user can be derived as
\begin{align}\label{passiveyu1}
    &\widetilde{y}_u^p=\sqrt{P_t}\big(\sqrt{L_fL_g}\sum_{n=1}^{N}|f(n)||p(n)||g(n)|+\sqrt{L_h}|h|\big)s+w_u.
\end{align}

\par The total power received by passive RIS is as follows
\begin{align}\label{passivePin}
&P^p_{in}=\sum_{n=1}^{N}{y_{in}^p}^\ast(n)y^p_{in}(n)=P_tL_g\sum_{n=1}^{N}|g(n)|^2.
\end{align}

\par Assuming that passive RIS uses the same reflection coefficient for each channel, according to \cite{ref5}, since passive RIS does not have signal amplification capability, $|p(n)|$ can be set to 1, the reflected power of passive RIS can be denoted as
\begin{align}\label{passivePout}
&P^p_{out}=\sum_{n=1}^{N}{y_{out}^p}^\ast(n)y^p_{out}(n)=P_tL_g\sum_{n=1}^{N}|g(n)|^2|p(n)|^2=P_tL_g\sum_{n=1}^{N}|g(n)|^2.
\end{align}

\par Therefore, the power consumption in the  system is given by
\begin{align}\label{passiveP_{tot}}
    &P^p_{tot}=\mu P_t+P_0+NP^p_{c,n}+P^p_{0,RIS},
\end{align}
\noindent where $P_{c,n}^p$ represents the static power required for reconfiguring each passive RIS element, while $P^p_{0,RIS}$ accounts for the remaining static power components in passive RIS system.

\par Similar to the derivation of active RIS, by using the law of large numbers, the SNR obtained by user can be converted into
\begin{align}
    &\gamma^p_u=\frac{ P_t\big(\sqrt{L_fL_g}\sum_{n=1}^{N}|f(n)||g(n)|+\sqrt{L_h}|h|\big)^2}{\sigma^2_u}\\
    &~~~=\frac{ P_t\big(\sqrt{L_fL_g}N\cdot\frac{1}{N}\sum_{n=1}^{N}|f(n)||g(n)|+\sqrt{L_h}|h|\big)^2}{\sigma^2_u}\nonumber\\
    &~~~\approx\frac{ P_t\big(\sqrt{L_fL_g}N\mathbb{E}\big\{|f(n)||g(n)|\big\}+\sqrt{L_h}|h|\big)^2}{\sigma^2_u}=\frac{ P_t\big(\frac{\pi}{2}\sqrt{L_fL_g}N\alpha_f\alpha_g+(\frac{\pi}{2})^\frac{1}{2}\sqrt{L_h}\alpha_h\big)^2}{\sigma^2_u}.\nonumber
\end{align}

\par Ultimately, for passive RIS-aided system, the EE is expressed as
\begin{align}
    \mathrm{EE}^p=B\cdot\frac{\mathrm{log}_2(1+\gamma^p_u)}{P^p_{tot}},
\end{align}
\noindent the EE can be further expressed as
\begin{align}
    \mathrm{EE}^p(N, P_t, \sigma^2_u)=B\cdot\frac{\mathrm{log}_2(1+\frac{P_t(A_8N^2+A_9N+A_{10}}{\sigma^2_u})}{P_t+NP^p_{c,n}+A_{11}},
\end{align}
\noindent where
\begin{align}
    &A_8=\frac{\pi^2}{4}L_fL_g\alpha^2_f\alpha^2_g,~~~~~A_9=\pi(\frac{\pi}{2})^{\frac{1}{2}}\sqrt{L_fL_gL_h}\alpha_f\alpha_g\alpha_h,\nonumber\\
    &A_{10}=\frac{\pi}{2}L_h\alpha^2_h,~~~~~~~~~~~A_{11}=P_0+P^p_{0,RIS}.
\end{align}

}


 

\begin{thebibliography}{1}
\bibliographystyle{IEEEtran}

\bibitem{part0_1}J. Li, L. Xu, P. Lu, et al., ``Performance analysis of directional modulation
with finite-quantized RF phase shifters in
analog beamforming structure," \textit{IEEE Access}, vol. 7, pp. 97457--97465, Jul. 2019.

\bibitem{part0_2}R. Dong, B. Shi, F. Shu et al., ``Performance analysis of massive hybrid directional modulation with mixed phase shifters,"  \textit{IEEE Trans. on Veh. 
Technol.}, vol. 71, no. 5, pp. 5604--5608, May. 2022.

\bibitem{ref1}R. Liu, K. Katsanos, Q. Wu, et al., ``Overview of IRS for 6G and industry advance," \textit{Intelligent Surfaces Empowered 6G Wireless Network}, pp. 83--115, 2024.

\bibitem{2}M. Alsabah, M. Naser, B. Mahmmod, et al., ``6G wireless communications networks: A comprehensive survey," \textit{IEEE Access}, vol. 9, 
 pp. 148191--148243, Jan. 2024.

\bibitem{3}Q. Wu, R. Zhang, ``Intelligent reflecting surface enhanced wireless network: Joint active and passive beamforming design," \textit{2018 IEEE Global Communications Conference}, pp. 1--6, Feb. 2018.

\bibitem{4}Q. Li, M. El-Hajjar, I. Hemadeh, et al., ``Reconfigurable intelligent surface aided amplitude and phase-modulated downlink transmission," \textit{IEEE Trans. on Veh. 
Technol.}, vol. 72, no. 6, pp. 8146--8151, Jun. 2023.

\bibitem{part1_1}C. Huang, A. Zappone, G. C. Alexandropoulos, et al., ``Reconfigurable intelligent surfaces for energy efficiency in wireless communication," \textit{IEEE T. Wirel. Commun.}, vol. 18, no. 8, pp. 4157--4170, Aug. 2019.

\bibitem{part1_2} Q. Wu and R. Zhang, ``Towards smart and reconfigurable environment:
Intelligent reflecting surface aided wireless network,” \textit{IEEE Commun. Mag.}, vol. 58, no. 1, pp. 106--112, Jan. 2020.

\bibitem{part2_4}Y. Gao et al., ``Reflection resource management for intelligent reflecting
surface aided wireless networks,” \textit{ IEEE Trans. on 
Commun.}, vol. 69, no. 10,
pp. 6971--6986, Oct. 2021.

\bibitem{part2_5}W. Tang et al., ``MIMO transmission through reconfigurable intelligent
surface: System design, analysis, and implementation,” \textit{IEEE J. Sel. Areas Commun.}, vol. 38, no. 11, pp. 2683--2699, Nov. 2020.

\bibitem{ref3}O. Dikmen, ``Performance analysis and simulation of IRS-aided wireless networks communication," \textit{Symmetry}, vol. 16, no. 2, Mar. 2024.

\bibitem{ref5}R. Dong, Y. Teng, Z. Sun, et al.,  ``Performance analysis of wireless network aided by discrete-phase-shifter IRS," \textit{J. Commun. Netw.}, vol. 24, no. 5,
pp. 603--612, Aug. 2022.

\bibitem{new5_31}L. Yang, J. Yang, W. Xie, et al., ``Secrecy Performance Analysis of RIS-Aided Wireless Communication Systems," \textit{IEEE Trans. on Veh. 
Technol.}, vol. 69, no. 10, pp. 12296--12300, Oct. 2020.

\bibitem{part2_15}R. Dong, Z. Xie, F. Shu, et al., ``Performance analysis of discrete-phase-shifter IRS-aided amplify-and-forward relay network,'' \textit{J. Election. Inf. Techn.},  vol. 47, no. 1, pp. 138--146, 2025.

\bibitem{ref6}H. Liu, N. Qi, K. Wang, et al., ``Network deployment with energy efficiency optimization in IRS-assisted cell-free MIMO networks," \textit{Physical Commun.}, vol. 63, Mar. 2024.

\bibitem{part2_17}F. Shu et al., ``Enhanced secrecy rate maximization for directional modulation networks via IRS,"  \textit{IEEE Trans. Commun.}, vol. 69, no. 12, pp. 8388--8401, Dec. 2021.

\bibitem{part2_11}H. Sarieddeen et al., ``Terahertz-band MIMO systems: Adaptive transmission and blind parameter estimation," \textit{IEEE Commun. Lett.}, vol. 25, no. 2, pp. 641--645, Feb. 2021.

\bibitem{part2_12}R. Liu, Q. Wu, M. Di Renzo, et al., ``A path to smart radio
environments: An industrial viewpoint on reconfigurable intelligent
surfaces,” \textit{IEEE Wireless Commun.}, vol. 29, no. 1, pp. 202--208,
2022.

\bibitem{part2_13}M. Najafi, V. Jamali, R. Schober, et al., ``Physics-based modeling and scalable optimization of large intelligent reflecting surfaces,”
\textit{IEEE Trans. Commun.}, vol. 69, no. 4, pp. 2673--2691, Apr. 2021.

\bibitem{part3_2}W. Lv, J. Bai, Q. Yan, et al., ``RIS-assisted green secure communications: Active RIS or passive RIS?," \textit{IEEE Wireless Commun. Lett.}, vol. 12, no. 2, pp. 237--241, Feb. 2023.

\bibitem{11}Z. Zhang, L. Dai, X. Chen, et al., ``Active RIS vs. passive RIS: Which will prevail in 6G?,'' \textit{IEEE Trans. Commun.}, vol. 71, no. 3, pp. 
1707--1725, Mar. 2023.

\bibitem{part3_7}W. Shi, Q. Wu, F. Shu, et al., ``Joint transmit and reflective beamforming design for active IRS-aided SWIPT systems,"  \textit{Chinese J. Electron.}, vol. 33, no. 2, pp. 536--548, Mar. 2024.

\bibitem{ref7}S. Tsai, and H. Poor,  ``Power allocation for artificial-noise
secure MIMO precoding systems," \textit{IEEE Trans. on Signal Process.}, vol. 62, no. 13, pp. 
3479--3493, Jun. 2014.

\bibitem{part3_4}Y. Zhao, X. Wang, F. Shu, et al., ``Joint power allocation and beamforming design for active IRS-aided secure directional modulation systems," \textit{IEEE Open J. Commun. Soc.}, vol. 6, pp. 2853--2865, Nov. 2025.

\bibitem{part3_6}F. Shu et al., ``Beamforming and transmit power design for intelligent reconfigurable surface-aided secure spatial modulation," \textit{IEEE J. Sel. Topics Signal Process}, vol. 16, no. 5, pp. 933--949, Aug. 2022.

\bibitem{ref10}Y. Wang, F. Shu, Z. Zhuang, et al., 
 ``Asymptotic performance analysis of large-scale active IRS-aided wireless network," \textit{IEEE Open J. Commun. Soc.}, vol. 4, pp. 2684--2696, Oct. 2023.

\bibitem{part3_0}K. Zhi, C. Pan, H. Ren, et al., ``Active RIS versus passive RIS: Which is superior with the same power budget?,"  \textit{IEEE Commun. Lett.}, vol. 26, no. 5, pp. 1150--1154, May. 2022.

\bibitem{part4_0}S. Jian, L. Jifeng, L. Xiao, et al., ``Energy-efficient encoding for RIS-assisted communication system under measurement-based power consumption: Method and field trials," \textit{China Communications}, vol. 22, no. 4, pp. 281--295, Apr. 2025.

\bibitem{part4_3}D. Jia, Y. Zhong, X. Zhou, et al., ``Energy efficiency in RIS-assisted wireless networks: Impact of phase shift and deployment," \textit{IEEE Wireless Communications and Networking Conference}, pp. 1--6, 2024.

\bibitem{part4_2}J. Chen, Z. Ma, Y. Zou, et al., ``DRL-based energy efficient resource allocation for STAR-RIS assisted coordinated Multi-cell networks,"  \textit{IEEE Global Communications Conference}, pp. 4232--4237, 2022.

\bibitem{powerconsumpation}R. K. Fotock, A. Zappone and M. D. Renzo, ``Energy efficiency optimization in RIS-aided wireless networks: Active versus nearly-passive RIS with global reflection constraints," \textit{IEEE Trans. Commun.}, vol. 72, no. 1, pp. 257--272, Jan. 2024.

\bibitem{dashudinglv}L. Wasserman,  ``All of statistics: A concise course in statistical inference,” \textit{New York, NY, USA: Springer}, 2004.

\bibitem{netown}Q. Cheng, R. Dong, W. Cai, et al., ``Two enhanced-rate power allocation strategies for active IRS-assisted wireless network," \textit{International Conference on Computer Communication and Artificial Intelligence}, pp. 458--463, 2024.

\bibitem{erfen}G. Walter, ``Numerical analysis,” Dec. 2013.

\bibitem{ref16}S.-Y. Jung, J. Hong, and K. Nam, ``Current minimizing torque control of
the IPMSM using Ferrari’s method,” \textit{IEEE Trans. Power Electron.}, vol. 28, no. 12, pp. 5603--5617, Dec. 2013.










\end{thebibliography}
%

\vfill

\end{document}